# Actively switchable non-degenerate polarization entangled photon pair distribution in a dense wave-division multiplexing

Zhi-Yuan Zhou[1+], Yun-Kun Jiang[2+#], Dong-Sheng Ding[1], Bao-Sen Shi[1*], Guang-Can Guo[1]

[1]*Key Laboratory of Quantum Information, University of Science and Technology of China, Hefei 230026, P. R. China*

[2]*College of Physics and Information Engineering, Fuzhou University, Fuzhou, 350002, P. R. China*

[+]These authors contribute equally to this work.

[*] *drshi@ustc.edu.cn*

[#] *ykjiang@fzu.edu.cn*

We have realized experimentally a non-degenerate polarization-entangled photon pair distribution in a commercial telecom dense wave-division multiplexing device (DWDM) with 8 channels. A promising point of this work is that the entangled photon pair is obtained via spontaneously parametric down conversion in a single type-II periodically poled $KTiOPO_4$ crystal without usual post-selection. Another promising advantage is that we can actively control and switch the distribution of the photon pair between different channel pairs in DWDM at will. There is no crosstalk between the different channel pair because of a limited emission bandwidth of the source. High raw visibility of greater than 90% in Bell-type interference measurement in each channel pair and the CHSH inequality S parameter of 2.63±0.08 prove high entanglement of our source. Our work is helpful for building quantum communication networks.

Dense wave-division multiplexing (DWDM) is a crucial technology used in classical optical communication networks, it can dramatically increase the transmission capacity of a single fiber communication channel. A combination of this technique with a photon source prepared by the spontaneously parametric down conversion (SPDC) can help one to distribute entangled non-degenerate photon pairs to a large number of users, which is a key step for building a quantum network. There are some reported works related to the combination of DWDM and photon pairs [1-5]. Jiang et al [1] used a coarse WDM to generate two non-degenerate polarization entangled photon pairs. Lim et al [2] performed a proof-of-principle experiment to separate the photon pairs to simulate the de-multiplexing operation with the aid of a dichroic mirror and tunable filters inserted on each channel. Moreover, a passive wavelength selective switch, based on arrayed waveguide grating technology, was tested in [3]. Most of the previous works use the photon with the spectral bandwidth of tens of nanometers [1-3, 5], and entanglement exists in channel pairs symmetric to the center of the emission spectrum. These schemes can realize the photon pair distribution in a channel pair passively using DWDM, but it is difficult to control and switch actively the photon pair distribution between different channel pairs. What we have done here is very different compared to the previous works. The photon pair is generated via SPDC in a single type-II periodically poled $KTiOPO_4$ (PPKTP) crystal, the emission spectral bandwidth of the photon is 2 nm, which is within the channel distance of a 200-GHz DWDM we use. The central wavelength of the generated photon can be continuously adjusted over a broad range by tuning the pump wavelength and the crystal temperature [6]. When the central wavelength of the emission photon is tuned to the center between two adjacent channels of DWDM, the outputs from these two channels are entangled, and a non-degenerate polarization entangled photon pair is obtained. Therefore we can realize the active control and switch of photon pair distribution among different adjacent channels at will by simply tuning the pump wavelength and the crystal temperature. We also experimentally find that there is no crosstalk between the different channel pairs. Besides, another promising point we want to show is that the entangled photon pair is obtained without usual postselection. This source is different from a traditional source obtained by post-selecting cross-sections of two rings [7]. Our source is also different from schemes shown in Refs, 1, 2, 5, 8-14, where either two nonlinear crystals [5, 8, 9] or an interferometer [1, 2, 10-14] configuration are needed, which are rather complex. Another point we want to mention is that although this source seems similar to scheme shown in Ref. 6, they are different. The source of Ref. 6 is based on the usual postselection, the output is the degenerate photon pair. On the contrary, the photon pair is obtained in determinacy in this work. We believe that this work is promising for building quantum communication networks

In the text following, the experimental setup for our DWDM photon distribution system is described firstly. Then we characterize the performance of different adjacent channel pairs using different kinds of measurement. The Bell-type interference curve is checked and the CHSH inequality S parameter value is calculated. By which, we could check the entanglement of our source. We also measure the brightness of different channel pairs. Factors that limit the performance of our system are discussed in text.

The layout of our source is depicted in figure 1. The pump light beam is from a continuous wave (CW) Ti: Sapphire laser (Coherent MBR110), its wavelength can be continuously tuned from 756 nm to 870 nm. The polarization of the pump beam is controlled by a half wave plate. The pump beam is focused into a 10-mm long type-II phase-matched PPKTP crystal by lens L1. The group delay between the down-converted signal and idle photons inside the PPKTP crystal is compensated by a 5-mm long KTP crystal with its optical axis rotated by 90 degree relative to the PPKTP crystal. The temperatures of both crystals are stabilized using two temperature controllers separately. After SPDC, the strong pump beam is filtered by a dichromatic mirror and a long-pass filter. After that, the signal and idler photons are collected using a single mode fiber (SMF). The collected photons pass through a commercial 200 GHz DWDM (120 GHz channel bandwidth and 80 GHz channel separation, the central wavelengths for channel 20 to 34 are 1561.25, 1559.75, 1558.00, 1556.35, 1554.7, 1553.15, 1551.55 and 1549.95 nm respectively measured with a accuracy of ±0.15 nm). A non-degenerate polarization entangled photon pair is obtained in two adjacent channels of DWDM output. Two fiber polarization rotators, polarizer (FPR&P) and one fiber arbitrary retarder (FAR) are used to characterize the quality of entanglement. The output of FPR&P1 is connected to a single-photon detector (InGaAs avalanche photodiode 1, APD1, Lightwave Princeton ) directly, and the output of FPR&P2 is optically delayed using a 200-m long SMF before connecting to APD2 (Qasky). APD1 runs at external trigger mode with a trigger rate of 66.6 MHz with 1 ns detection window and a detection efficiency of 15%, the dark counts rate is $6\times10^{-6}ns^{-1}$. The output of APD1 is electrically delayed using a digital delay generator (DG535, Stanford Instr.). The electrical signal from DG535 is used to trigger APD2. APD2 has a detection efficiency of 8% and detection window of 2.5 ns, the dark counts is $6.5\times10^{-6}$ ns$^{-1}$. The output of APD2 is connected to a counter to accumulate the coincidence counts.

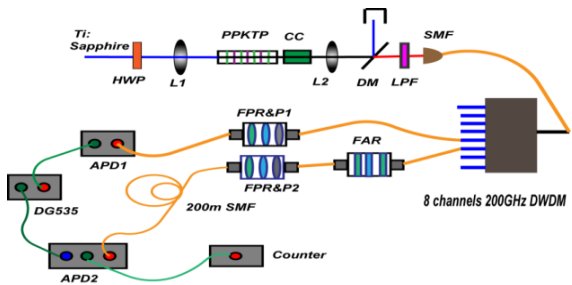

**Figure** 1. The layout of the experiment. HWP: half wave plate; L1, L2: lens; PPKTP: periodically-poled KTiOPO4 crystal; CC: KTP compensate crystal; DM: dichromatic mirror; LPF: long pass filter; SMF: single mode fiber; FAR: fiber arbitrary retarder; FPR&P1, 2: fiber polarization rotator and polarizer; APD1, 2: InGaAs avalanche photodiode; DG535: digital delay generator.

The present source bases on our previous developed broadband continuously tunable source [6]. The bandwidth of the down-converted photon is 2 nm (246 GHz). By tuning the pump wavelength and the temperature of the PPKTP crystal, we can make the emission spectrums of the two down-converted orthogonal photons overlap, then a suitable spectrum selection is used to prepare a non-degenerate polarization entangled photon pair. The operation principle of our source is depicted in figure 2. The central wavelength of the down-converted photon pair is tuned to the center of two adjacent channels of DWDM (red solid curve), two adjacent channels of DWDM are turned to select about 120 GHz wide spectrum symmetric to the center of the photon spectrum, then a non-degenerate polarization entangled photon pair can be directly obtained from the outputs of this pair channels. The output state of the photon pair from the DWDM can be expressed as

$$|\Phi\rangle = \frac{1}{\sqrt{2}}\left(|H(\lambda_s)V(\lambda_i)\rangle + e^{i\varphi}|V(\lambda_s)H(\lambda_i)\rangle\right) \quad .(1)$$

Where H and V stand for horizontal and vertical polarizations, $\lambda_s$ and $\lambda_i$ are wavelengths for signal and idler photons, $\varphi$ is a relative phase between the two polarizations due to phase mismatch arising from the different wavelength of signal and idler, and this phase can be compensated using a fiber arbitrary retarder. For actively switching the photon pair to other adjacent channel pairs, we just need to move the center of the photon emission spectral to lie in the center of them.

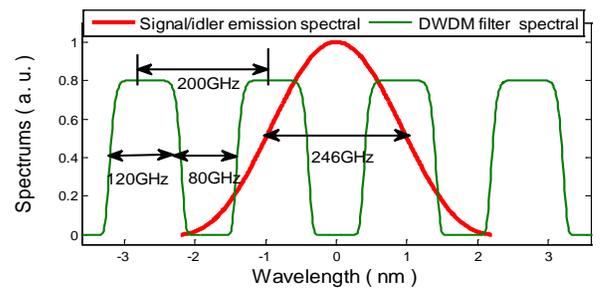

**Figure 2.** Operation principle of our experiment. The red solid curve is the emission spectral of signal (idler), the green curve is filter spectral of 200 GHz DWDM.

Different types of measurements are used to characterize the quality of entanglement. In the experiment, Bell-type interference is measured between channel 22 and channel 20. Channel 22 is connected to APD1 and channel 20 is connected to APD2. The pump power is 204 mW and the temperature of the PPKTP crystal is 23.4$^0$C. The single count of APD1 is about 9000 per second and the dark count of APD1 is about 400/s, the average coincidences is 34s$^{-1}$, the accidental coincidences is about 0.4s$^{-1}$, which can be neglected. The results are showed in figure 3(a), raw visibilities at 0/90$^0$ and +/-45$^0$ bases are (96.05±0.02)% and (90.77±0.07)% respectively. Such high visibility (>71%) is sufficient for the violation of Bell-inequality, and implies a high polarization entanglement of our source. The interference curves between other channel pairs are not shown here, instead we give the raw visibilities in the two bases shown in figure 3(b), we can see that all the raw visibilities are equal or greater than 90%. The pump power is kept at 204 mW when we tune the wavelength of the pump to make the central wavelength of the down-converted photon pairs been in the center of a certain channel pair. The average coincidences for different channel pairs are showed in figure 3(c), the difference in coincidences rate is mainly due to non-uniformity of channel bandwidth (see the central wavelength for different channel shown above) and a wavelength dependent detection efficiencies of APD1 and ADP2.

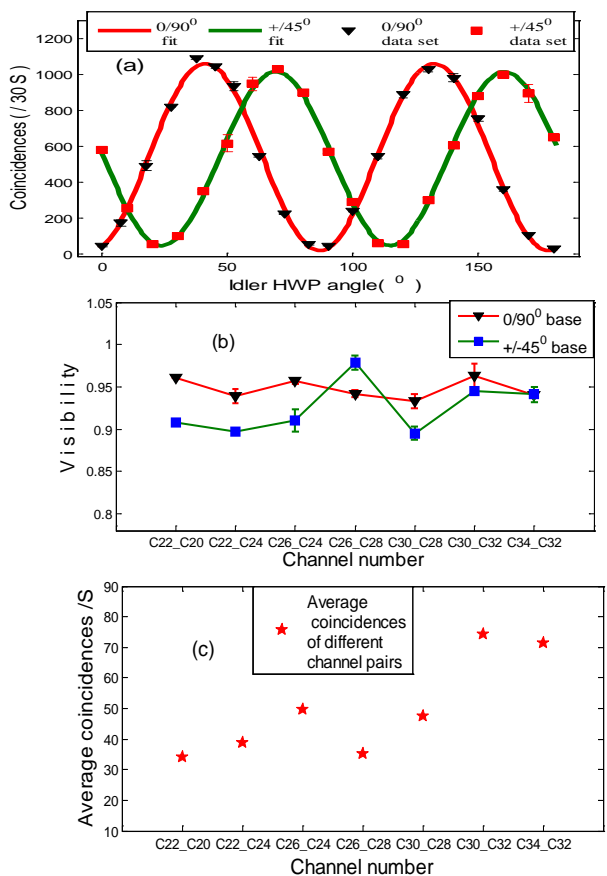

Figure 3. (a)Coincidences per 30 seconds as function of the half wave plate angle inserted in idler beam, the angle of half wave plate inserted in signal beam is keep at $0^0$ or $45^0$;(b) Visibilities at $0/90^0$ and $+/-45^0$ bases for different channel pairs. The horizontal axis is channel pairs of successive channels, the order of the channels means that the former channel is connected to APD1 and the detection output of ADP1 is used to trigger APD2 connected with the latter channel. (c) Average coincidences per seconds for different channel pairs.

The experimental conditions for data measurements of different channel pairs are list in table 1.The difference in time delay is a result of the add-drop filter mechanism of DWDM. The difference in single count rate is due to the non-uniformity of channel bandwidth and inaccuracy of the center wavelength.

Table 1. Experimental parameters for different channel pairs

| Channel numbers | Pair central Wavelength (nm) | Delay (ns) | single counts(s$^{-1}$) | Temperature ($^0$C) |
|---|---|---|---|---|
| C22_C20 | 1560.50 | 804.4 | 9000 | 23.40 |
| C22_C24 | 1558.88 | 794.8 | 8000 | 25.13 |
| C26_C24 | 1557.18 | 804.2 | 9000 | 27.10 |
| C26_C28 | 1555.53 | 794.5 | 7500 | 29.00 |
| C30_C28 | 1553.93 | 804.2 | 9000 | 31.28 |
| C30_C32 | 1552.35 | 793.9 | 12000 | 33.72 |
| C34_C32 | 1550.55 | 804.2 | 12000 | 36.20 |

In order to demonstrate further the entanglement between the photons in a pair, we check the CHSH inequality. It is well known that CHSH inequality $|S| \leq 2$ holds if there is no entanglement between photons in a pair. On the contrary, it is violated if there is entanglement. Where S is defined as [7]

$$S = E(\theta_1, \theta_2) + E(\theta_1', \theta_2) + E(\theta_1, \theta_2') + E(\theta_1', \theta_2') \quad (2)$$

And $E(\theta_1, \theta_2)$ is given by

$$\frac{C(\theta_1, \theta_2) + C(\theta_1^\perp, \theta_2^\perp) - C(\theta_1, \theta_2^\perp) - C(\theta_1^\perp, \theta_2)}{C(\theta_1, \theta_2) + C(\theta_1^\perp, \theta_2^\perp) + C(\theta_1, \theta_2^\perp) + C(\theta_1^\perp, \theta_2)} \quad (3)$$

Where $C(\theta_1, \theta_2)$ is coincidence count at different settings of the two polarizers. The settings we used in our experiments are $\theta_1 = -22.5^0$, $\theta_1^\perp = 67.5^0$; $\theta_1' = 22.5^0$, $\theta_1'^\perp = 122.5^0$; $\theta_2 = -45^0$, $\theta_2^\perp = 45^0$; $\theta_2' = 0^0$, $\theta_2'^\perp = 90^0$. We calculate the S parameter of the CHSH inequality using the data we measure between channel 32 and 34, which is 2.63±0.08, a violation of CHSH inequality with 8 standard deviations.

The photon pairs experience some loss optical elements before detected by single photon detectors. The main losses are filtering loss (1.67dB), single mode fiber coupling loss (4.68dB), insertion loss of DWDM (0.76dB) and FPR&P1 (0.86dB), FPR&P2 (1.16dB) and FAR (1.05dB). Counting for all these loss factors, the total losses for signal and idler photon are 7.97dB and 9.32dB respectively. Given the detection efficiencies and detection window of the single photon detectors, we estimate an average photon pair generation rate of 7.4×10$^3$ (mW·nm)$^{-1}$. We use $R_{estimate} = R_{detected}/(\alpha_s \alpha_i \eta_1 \eta_2 d)$ for estimation, where $R_{detected} = 45$ is the average detected coincidences, $\alpha_{s,i}$ is the total loss for signal (idler) photon, $\eta_{1,2}$ is the detection efficiency of single photon detectors, and $d = 6.6\%$ is the detection duty cycle of APD1. The generation rate could be increase by for example selecting a DWDM with wider channel width and smaller channel separation, or using a PPKTP waveguide. We also experimentally check whether there is cross-talk between different channel pairs, our experimental results clearly show no, this is because the bandwidth of the photon emission spectral is less than the two sequential channels.

In conclusion. In this work, we experimentally generate a non-degenerate polarization entangled photon pair through SPDC in a type-II phase-matched PPKTP crystal in determinacy, and demonstrated experimentally that the photon pair could distribute in a channel pair of a commercial DWDM system. We can select a channel pair as output channels at will by simply tuning the pump wavelength and the crystal temperature. High visibilities and violation of CHSH inequality demonstrate high quality of our source. There is no crosstalk between different channel pairs due to a small emission spectral of our source. The results demonstrated in this article will be helpful for building future quantum communication network.

**Acknowledgements**

We thank Dr. Wei Chen and Dr. Bi-Heng Liu for kindly loan us single photon detector and for other technique support. This work was supported by the National Natural Science Foundation of China (Grant Nos., 11174271, 61275115), the National Fundamental Research Program of China (Grant No. 2011CB00200), and the Innovation fund from CAS, Program for NCET.